\documentclass[a4paper]{jpconf}
\usepackage{graphicx}
\usepackage{amsmath,amssymb}

\usepackage{braket}
\usepackage{cite}
\def\squarebr#1{\left[#1\right]}
\def\Ob{\mathcal{O}}
\def\n{\mathbf{n}}
\newcommand{\I}{\mathrm{i}}
\newcommand{\D}{\mathrm{d}}
\newcommand{\eq}[1]{\begin{equation}\begin{aligned}#1\end{aligned}\end{equation}}
\def\del#1#2{\frac{\partial #1}{\partial #2}}
\def\Del#1#2{\frac{\D #1}{\D #2}}
\def\eref#1{(\ref{eq:#1})}
\def\K{\mathbf{k}}
\def\W{\mathrm{W}}

\def\Hs{\mathsf{h}}
\begin{document}
\title{Beyond semiclassical time: dynamics in quantum cosmology}

\author{Leonardo Chataignier$^{1,2}$}

\address{$^1$Dipartimento di Fisica e Astronomia, Universit\`{a} di Bologna,
via Irnerio 46, 40126 Bologna, Italy}
\address{$^2$I.N.F.N., Sezione di Bologna, I.S. FLAG, viale B. Pichat 6/2, 40127 Bologna, Italy}

\ead{leonardo.chataignier@unibo.it}

\begin{abstract}
We review two approaches to the definition of the Hilbert space and evolution in mechanical theories with local time-reparametrization invariance, which are often used as toy models of quantum gravity. The first approach is based on the definition of invariant relational observables, whereas the second formalism consists of a perturbative construction of the Hilbert space and a weak-coupling expansion of the Hamiltonian constraint, which is frequently performed as part of the Born-Oppenheimer treatment in quantum cosmology. We discuss in which sense both approaches exhibit an inner product that is gauge-fixed via an operator version of the usual Faddeev-Popov procedure, and, in the second approach, how the unitarity of the effective Schr\"{o}dinger evolution is established perturbatively. We note that a conditional probability interpretation of the physical states is possible, so that both formalisms are examples of quantum mechanics with a relational dynamics.
\end{abstract}

\section{Introduction}
In any approach to quantum gravity, a central issue is the combination of spacetime diffeomorphism invariance with the standard postulates of quantum mechanics. Indeed, our current accepted formulation of quantum theory refers to a fixed spacetime background and it seems to leave little room for local time reparametrization invariance, i.e., for formulating the theory [especially the wave function(al)'s probabilistic interpretation and the state update rule] without a preferred, global class of time variables. In general relativity (GR), on the contrary, spacetime diffeomorphism invariance is an important ingredient, one that is intertwined with the delicate issue of background independence (see \cite{background-indep,relobs-2} for further discussion on this point) and, in particular, no preferred or global time seems to be generally available in GR. If gravity is to be quantized, one is thus faced with the choice of either modifying relativity to accommodate a preferred class of time variables, or formulating quantum theory without a fixed spacetime background. The latter choice is usually preferred and taken as a template for quantum gravity.

A background-independent quantization of gravitation faces serious challenges, as one needs to understand how to formulate the dynamics in a diffeomorphism-invariant fashion (or, at very least, explain the emergence of this diffeomorphism symmetry at the classical level). Can we maintain quantum theory's linear structure (the superposition principle) in this way? What is the Hilbert space? What are the observables? Most importantly, how can we assign a probabilistic interpretation to the wave function(al) without a preferred class of times with respect to which probabilities are conserved? With respect to which time (if any) should the state update rule be applied? Does the measurement problem become even more exacting in this case? In fact, what is the meaning of probabilities if the theory is extrapolated to the Universe, as is done in quantum cosmology?

In this conference article, we review two approaches that may be seen as templates to a version of quantum mechanics in which: no preferred time is assumed a priori; a notion of quantum diffeomorphism invariance holds; the linearity of the theory is maintained and the Hilbert space structure can be constructed, at the very least, perturbatively; a notion of unitarity and time evolution are available. In this way, the formalisms discussed here could be a starting point in our endeavour to address the above questions. Our procedures draw an analogy to well-known techniques in classical gauge systems, and we deal with mechanical models to avoid topics of field theory that are inessential to the problem at hand (quantum time diffeomorphism invariance and unitary time evolution). These models are directly applicable to homogeneous (classical and quantum) cosmology, and also to the case in which the usual Mukhanov-Sasaki variables are present \cite{MS,MS-1}.

In Sec. \ref{sec:gauge}, we briefly discuss how the classical diffeomorphism symmetry of GR can be seen as a type of gauge symmetry, and how this affects the definition of observables in the theory. In particular, in Sec. \ref{sec:worldline}, we restrict ourselves to the mechanical case, which is then quantized in Sec. \ref{sec:q-worldline}. A general, model-independent definition of gauge-invariant observables for theories with local time reparametrization invariance is proposed in Sec. \ref{sec:qrelobs}. Furthermore, we discuss a perturbative construction of the classical Hamilton-Jacobi solutions, as well as the quantum Hilbert space in Sec. \ref{sec:weak-coupling}, where the perturbative unitarity of the theory is also ascertained in a weak-coupling expansion. We conclude in Sec. \ref{sec:conclusions}.

\section{\label{sec:gauge}Gauge symmetry and observables}
\subsection{The classical theory}
In classical GR, diffeomorphisms are a local symmetry of the Einstein field equations. This means that, given a solution $\Phi$ (which denotes a set of tensor fields that includes the spacetime metric) and a diffeomorphism $\phi:\mathcal{M}\to\mathcal{M}$ (where $\mathcal{M}$ is the spacetime manifold), then the transformed fields $\phi\cdot\Phi$ also constitute a solution to the same field equations \cite{background-indep}. It is for this reason that GR is a diffeomorphism-invariant theory.

A local symmetry of the field equations implies a type of indeterminism, which can be discerned in the initial-value problem formulation of the theory \cite{background-indep}. Indeed, let $\mathcal{C}$ be a Cauchy slice on which the initial data are prescribed. If the diffeomorphism $\phi$ coincides with the identity in a neighborhood of $\mathcal{C}$ but is nontrivial elsewhere, then $\Phi$ and $\phi\cdot\Phi$ are to two different solutions associated with the same initial conditions. To restore the one-to-one correspondence between solutions and initial data, one is led to the notion of equivalence classes of $\Phi$ under diffeomorphisms, and indeed to the proposal that observables (here understood as quantities that have a deterministic evolution) should be invariant under diffeomorphisms. An analogous situation occurs in Yang-Mills theories, where the local symmetry transformations are not diffeomorphisms but gauge transformations. In that case, one is led to the notion of gauge-invariant quantities (e.g., an electric field) in order to overcome the ``gauge indeterminism''. Due to this similarity, we refer to any local symmetry as a ``gauge symmetry'' from now on. In this terminology, GR is also a gauge theory, albeit not of the Yang-Mills type.\footnote{There are efforts to construct a gravitational theory that follows more closely the Yang-Mills paradigm. In particular, the reader is referred to Poincar\'{e} gauge theory \cite{Poincare}.}

In general, diffeomorphism invariants (such as constant spacetime scalars) lack a clear physical meaning. Which of these invariants have a reasonable physical interpretation that can be related to experiments? The so-called relational observables \cite{relobs-2,relobs-0,relobs-1,relobs-11} (sometimes also referred to as ``evolving constants of motion'' \cite{evol-constants}) are invariants that have a straightforward interpretation. The idea is that observers record the dynamics of fields in local regions of spacetime in a relational way: one chooses a set of reference fields $\chi$ with respect to which the values of all dynamical fields are compared. Clearly, these reference fields are physical fields that are part of a solution $\Phi$ to the field equations. Ideally, the $\chi$ fields mark the passage of time and determine points in space, and thus they serve as ``generalized clocks and rods'' from which ``generalized referece frames'' can be specified. The values of all of the fields $\Phi$, taken relative to the clocks and rods, are the outcomes of experiments performed by the observers. These outcomes are the relational observables, which we denote by $\Ob[\Phi|\chi]$. But why are these objects invariant under diffeomorphisms?

In specifying the outcome $\Ob[\Phi|\chi]$ of an experiment (e.g., the value of an electric field relative to the values of four scalar fields that define a spacetime point \cite{gps}), one only measures physical fields. There is no need to make reference to arbitrary coordinate systems, nor to an abstract notion of a spacetime point $p\in\mathcal{M}$. For this reason, for any given value of the clocks and rods (e.g., $\chi=s$), the observable $\Ob[\Phi|\chi]$ does not depend on $p$, and it can be formally seen as a constant spacetime scalar, and thus a diffeomorphism invariant. More concretely, one can evidently choose $\chi$ to define an ``intrinsic coordinate system'' (i.e., one defined from the physical degrees of freedom instead of arbitrary constructs \cite{intrinsic}) in the abstract spacetime manifold $\mathcal{M}$, which corresponds to gauge fixing the diffeomorphism symmetry. Then, $\Ob[\Phi|\chi]$ are simply gauge invariant extensions of gauge-fixed tensor fields (see \cite{relobs-2} for further details on this point, and \cite{relobs-0} for the general notion of gauge invariant extensions).

Moreover, as the outcome $\Ob[\Phi|\chi]$ only refers to physical fields, it completely encodes the dynamics (in a relational way) in the region where the experiment is performed. The observables $\Ob[\Phi|\chi]$, when computed in theory, can also be seen as conditional quantities because they yield predictions for the observed outcomes given the condition that the reference fields have a certain value (the observed value of the clocks and rods).

\subsection{\label{sec:worldline}Worldline relational observables}
Let us discuss and illustrate the construction of relational observables in a simplified, mechanical setting. This is the same as considering physics in a $(0+1)$-dimensional spacetime or a worldline. In this case, a diffeomorphism-invariant theory for a set of scalar fields denoted by $q(\tau)$ is given by the action
\begin{equation}\label{eq:action}
S = \int_a^b\ \D\tau\ \mathcal{L}(q,\dot{q};e) = \int_a^b\ \D\tau\ \mathcal{L}(\phi^*q,\phi^*\dot{q};\phi^*e) \ ,
\end{equation}
where $\tau$ is an arbitrary coordinate on the worldline $\mathcal{M}$, $e(\tau)$ is the `einbein' in $\mathcal{M}$ {\cite{relobs-2}}, and $\phi$ is an arbitrary diffeomorphism on $\mathcal{M}$. Equation~\eref{action} implies that $S$ is gauge invariant because diffeomorphisms simply relabel the arguments of the Lagrangian, which retains its form under these transformations. More precisely, the action is exactly invariant if we restrict ourselves to diffeomorphisms that coincide with the identity at the endpoints $\tau = a, \tau = b$. For more general transformations, it is possible to add a boundary term to make the action fully gauge invariant {\cite{boundary}}. The invariance of the form of the action under (active) diffeomorphisms also implies that the Lagrangian $\mathcal{L}$ must satisfy
\eq{
\mathcal{L}\left(q'(\tau'),\Del{q'}{\tau'}(\tau');e'(\tau')\right) = \mathcal{L}\left(q(\tau),\Del{\tau}{\tau'}\Del{q}{\tau}(\tau);\Del{\tau}{\tau'}e(\tau)\right) = \Del{\tau}{\tau'}\mathcal{L}\left(q(\tau),\Del{q}{\tau}(\tau);e(\tau)\right)
}
under (passive) $\tau$ reparametrizations. This means that $\mathcal{L}$ is a homogeneous function of $\dot{q}$ and $e$, and therefore we must have
\eq{
\mathcal{L}(q,\dot{q};e) = \del{\mathcal{L}}{\dot{q}^i}\dot{q}^i+\del{\mathcal{L}}{e}e= p_i\dot{q}^i-e C(q,p) \ ,
}
where $p$ are the canonical momenta conjugate to $q$, and $C(q,p)$ is the first-class Hamiltonian constraint. Indeed, the equation of motion of the einbein is $C = 0$. Given the Poisson bracket structure for the $(q,p)$ pairs, one can verify that a gauge transformation (infinitesimal diffeomorphism) of the scalars reads $\delta_{\epsilon(\tau)}q(\tau) = \epsilon\dot{q} = \epsilon\{q,eC\}_{C=0}= \{q, \epsilon e C\}_{C=0} = \epsilon e \{q,C\}_{C=0}$ (and similarly for $p$). Thus, an invariant scalar $\Ob(q,p)$ Poisson-commutes with the constraint, $\{\Ob,C\}_{C=0}=0$. In contrast, a phase-space function $\eta(q,p)$ that is conjugate to the constraint, $\{\eta,C\}_{C=0}=1$, is a canonical representation of ``proper time'' because its equation of motion is $\dot{\eta}(\tau) = e(\tau)$, whereas ``proper time'' is the reparametrization invariant defined by $\eta := \int_a^b\D\tau\ e(\tau)$.

To find invariants that are physically meaningful, one can choose a reference field $\chi(\tau)$ to be the ``generalized clock'' in the worldline. This is a choice of gauge, and it is fixed by finding a diffeomorphism for which the pullback of $\chi(\tau)$ is trivial, $\phi^*\chi = \tau$. If this can be done, one can define the relational observable of any tensor field $f$ relative to $\chi$ as the constant worldline scalar
\eq{\label{eq:constant-relobs}
\Ob[f|\chi = s]:\, &\mathcal{M}\to\mathbb{R}\\
&p\mapsto \Ob[f|\chi = s](p) = \phi^*f \ (\forall p) \ ,
}
for each fixed value $\chi = s$ of the reference field. Each label $s$ corresponds to an instant. Thus, there is a family of constant scalars labelled by $s$, and this is why these objects are sometimes called ``evolving constants of motion'' {\cite{evol-constants}}. Despite their conceptual simplicity, it is often difficult to build relational observables in a model-independent way (i.e., without solving the equations of motion first). One useful definition considered in {\cite{relobs-11,relobs-2,int-relobs}} consists in reexpressing the pullback of $f$ in \eref{constant-relobs} as a spacetime (worldline) integral,
\eq{\label{eq:relobs-int}
\Ob[f|\chi = s] = \Delta_{\chi}\int_{\mathcal{I}}\D\tau\ \delta(\chi(\tau) - s)f(\tau) = \frac{\int_{\mathcal{I}}\D\tau\ \delta(\chi(\tau) - s)f(\tau)}{\int_{\mathcal{I}}\D\tau\ \delta(\chi(\tau) - s)}\ ,
}
where $\mathcal{I}$ is an interval of $\mathcal{M}$ where $\chi(\tau)$ is monotonic, $\Delta_{\chi}^{-1} = \int_{\mathcal{I}}\D\tau\ \delta(\chi(\tau)-s)$, and $\Delta_{\chi}=\phi^*\{\chi,C\}$ is the `Faddeev-Popov (FP) determinant' associated with the gauge choice $\phi^*\chi=\tau$ {\cite{relobs-11,relobs-2,FP}}. Given~\eref{relobs-int}, one can explicitly compute {\cite{relobs-2}} the gauge transformation $\delta_{\epsilon(\tau)}{\Ob[f|\chi = s]}_{C = 0} = \epsilon(\tau)e\{{\Ob[f|\chi = s]},C\}_{C = 0}=0$; i.e., one finds that the relational observables are indeed diffeomorphism invariant.

\subsection{\label{sec:q-worldline}Quantum theory. The physical Hilbert space}
We now present a possible construction of the quantum theory of a diffeomorphism-invariant mechanical model. The reader is referred to \cite{relobs-1,int-relobs,hsl} for other approaches, and to \cite{relobs-11,relobs-2} for more details concerning the formalism that we present here, the advantage of which is its model independence; i.e., the observables can be defined, in principle, without first solving the (classical) equations of motion.

The first step is to define an auxiliary Hilbert space $\mathcal{H}_{\text{aux}}$ on which the canonical quantization of the Hamiltonian acts as a self-adjoint operator $\hat{C}$ with a complete orthonormal system of eigenstates $\ket{E,\mathbf{k}}$. The label $E$ can be discrete or continuous and $\K$ stands for degeneracies. We will denote sums or integrals over $E$ by the formal symbol $\sum_E$. From these energy eigenstates, one can construct the `proper-time' states
\eq{\label{eq:proper-time-states}
\ket{t,\K} \propto \frac{1}{\sqrt{2\pi\hbar}}\sum_{E}\e^{-\frac\I\hbar E t}\ket{E,\K} \ , 
}
which are not necessarily orthonormal with respect to the auxiliary inner product in $\mathcal{H}_{\text{aux}}$ (see \cite{hsl,time-states}). This is, however, irrelevant for us because only physical states and invariant observables are meaningful, not the auxiliary structure of $\mathcal{H}_{\text{aux}}$. Indeed, the usefulness of the states \eref{proper-time-states} will be in the definition of a physical inner product, as we will see shortly.

In this theory, physical states  $\ket{\Psi}$ are superpositions of $\ket{E = 0,\K}$ so as to satisfy the quantum constraint $\hat{C}\ket{\Psi} = 0$, which is the celebrated Wheeler-DeWitt (WDW) equation {\cite{WDW}}. This condition implies diffeomorphism-invariance, since $\delta_{\epsilon(\tau)}\ket{\Psi} = \epsilon(\tau)\D\ket{\Psi}/\D\tau = \epsilon(\tau)\hat{C}\ket{\Psi}=0$. To define a Hilbert space from these physical states, it is necessary to define an inner product with respect to which $\ket{E = 0,\K}$ are, at the very least, delta-function normalizable. The auxiliary inner product reads $\braket{E = 0,\K'|E=0,\K} = \delta(0,0)\delta(\K',\K)$. If the zero energy eigenvalue is in the continuous spectrum, then $\delta(0,0) = \infty$.

To see how the inner product can be regularized, or how the $\delta(0,0)$ factor can be discarded, let us define from \eref{proper-time-states} the operator $\hat{\mu}_t := \sum_{\K}\ket{t,\K}\bra{t,\K}$, which satisfies $\int\D t\ \hat{\mu}_t= \hat{1}$. Then we may write
\eq{
\braket{E = 0, \K'|E=0, \K} &= \left(\int\D t\right) \left<E = 0, \K'\left|\hat{\mu}_t\right|E=0, \K\right>\\
&=\mathcal{V}\delta(\K', \K) =: \mathcal{V}(E = 0,\K'|E = 0,\K) \ .
}
The possibly divergent factor is clearly the ``volume'' of proper-time integration, $\mathcal{V}:=\int\D t$. The physical inner product is obtained by factoring out $\mathcal{V}$, and it reads
\eq{\label{eq:phys-IP}
(E = 0,\K'|E = 0,\K) := \left<E = 0, \K'\left|\hat{\mu}_t\right|E=0, \K\right> = \delta(\K', \K) \ .
}
This regularization of the inner product is analogous to the usual FP gauge fixing that is performed in path integrals, $\mathcal{Z} = \int\mathcal{D}\Phi\ \e^{\frac\I\hbar S} = \mathcal{V}\int\mathcal{D}\Phi\  \Delta_{\chi}\delta(\chi-s)\e^{\frac\I\hbar S}$, where $\hat{\mu}_t$ is analogous to the gauge-fixing delta function $\delta(\chi-s)$ for the proper-time gauge (for which $\Delta_{\chi}=\phi^*\{\chi,C\}=1$). The inner product~\eref{phys-IP} is also obtained by the Rieffel induction procedure \cite{Rieffel}. The physical Hilbert space is then defined to be the space of solutions of the WDW equation that are normalizable with respect to~\eref{phys-IP}.

\subsection{Quantum relational observables\label{sec:qrelobs}}
The physical inner product~\eref{phys-IP} also motivates a relevant definition of observables. General observables should be invariant symmetric operators that have the correct Heisenberg equations of motion. Here, we simply quote the definition (see {\cite{relobs-2}} for details)
\eq{\label{eq:qrelobs}
\hat{\Ob}[f|\chi = s] &:= \pi\hbar\sum_E\hat{P}_E[\hat{f}(\tau),\hat{\mu}_{t = s}]_+\hat{P}_E \ , 
}
where $\hat{P}_E$ is the projector onto the eigenspace associated with the energy $E$, $[\cdot,\cdot]_+$ is an anticommutator, and $\hat{f}(\tau)$ is in the Heisenberg picture. One can easily verify that~\eref{qrelobs} is an invariant in the sense that it commutes with $\hat{C}$, and we propose that~\eref{qrelobs} should be the general, model-independent definition of quantum relational observables (in the proper-time gauge) because:
\begin{enumerate}
\item It can be shown that~\eref{qrelobs} reduces to an integral formula analogous to~\eref{relobs-int}. For instance, if the spectrum of $\hat{C}$ spans the real line, than $\hat{\Ob}[f|\chi = s]=\frac12 \int_{-\infty}^{\infty}\D\tau\ \hat{f}(\tau)\hat{P}_{t = s-\tau}+\text{h.c.}$ (see {\cite{relobs-2}} for further details).
\item By construction, $\hat{\Ob}[1|\chi = s] = \hat{1}$ (FP resolution of the identity).
\item In general, the observables~\eref{qrelobs} satisfy the correct Heisenberg equations of motion, 
\eq{\label{eq:qrelobs-d}
\I\hbar\Del{}{s}\hat{\Ob}[f|\chi = s] = \hat{\Ob}\left[\left.\I\hbar\del{f}{s}+[f,C]\right|\chi = s\right] \ .
}
These are the correct equations because they are also satisfied by the classical observables if we replace $[\cdot,\cdot]/(\I\hbar)\to\{\cdot,\cdot\}$. In particular, notice that, due to~\eref{qrelobs-d}, dynamics need not  ``disappear'' in the quantum theory, as is often declared with regards to the WDW equation. Moreover, the evolution is with respect to a proper-time parameter that is defined directly from energy spectrum [cf.~\eref{proper-time-states}], and, in this sense, it is also an intrinsic evolution, since no external or preferred time is assumed a priori.
\end{enumerate}
It is also worth mentioning that, similarly to the conditional interpretation of the classical observables, the statistics of the quantum observables~\eref{qrelobs} can be seen as a conditional one. In {\cite{hsl}}, {a specific class of constraint operators and} another definition of relational observables was examined, and it was pointed out that the quantum dynamics of relational observables is equivalent to the one obtained in the famous Page-Wootters formalism {\cite{PW}}, which uses conditional probabilities. Thus, it is also instructive to note that the model-independent definition~\eref{qrelobs} leads to conditional expectation values
\eq{\label{eq:relobs-cond}
\braket{\hat{\Ob}[f|\chi = s]}_{\Psi} = \frac{\left(\Psi\left|\hat{\Ob}[f|\chi = s]\right|\Psi\right)}{(\Psi|\Psi)} = \frac{\braket{\Psi|\hat{f}\hat{\mu}_{t = s}|\Psi}}{\braket{\Psi|\hat{\mu}_{t = s}|\Psi}}=:\mathrm{E}_{\Psi}[f|\chi = s] \ .
}
This is also true for more general definitions of the gauge-fixing given by $\hat{\mu}_{t=s}$ (for example, for gauges with more than one frequency sector), as was shown in \cite{relobs-11,relobs-2}.

\subsection{A suggestion for the theory's postulates\label{sec:postulates}}
Equipped with a definition of the physical Hilbert space and observables, we now suggest a set of postulates for the theory, which can be seen as a conservative union of standard quantum mechanics and worldline diffeomorphism invariance. The reader is again referred to {\cite{relobs-2}} for a more detailed discussion.
\begin{enumerate}
\item In a quantum system with diffeomorphism invariance, the quantum state is a ray in the physical Hilbert space equipped with the inner product~\eref{phys-IP}.
\item A measurement of a tensor field $\hat{f}(\tau)$ has an eigenvalue $f(s,\mathbf{n})$ of a self-adjoint observable $\hat{\Ob}[f|\chi = s]$ as the outcome, the probability of which is
\eq{
p_{\Psi} = \frac{|(\n;s|\Psi)|^2}{(\Psi|\Psi)} \ ,
}
where we denote the eigenstates of $\hat{\Ob}[f|\chi = s]$ by $\ket{\n;s}$.
\item The state of the system is updated to $\ket{\n;s}$ after the measurement.
\end{enumerate}

\section{\label{sec:weak-coupling}Weak-coupling expansion}
The formalism described thus far leads to a model-independent and self-contained approach. Nevertheless, it is predicated on the assumption that it is possible to find the spectrum of $\hat{C}$ and its eigenstates in an exact, analytical manner [cf.~\eref{proper-time-states} and~\eref{qrelobs}]. What if we cannot solve the constraint? How do we construct the observables and the Hilbert space?

In this section, we consider an alternative, perturbative procedure that is directly applicable to toy models of the early Universe \cite{MS}, and that should be useful to more realistic models, for which it may be difficult or impossible to find the spectrum of $\hat{C}$ exactly. See {\cite{MS-1,relobs-2}} for further details. Henceforth, we set $\hbar=1$.

\subsection{Hamilton-Jacobi theory}
We consider the important case in which the Hamilton-Jacobi (HJ) formulation of the Hamiltonian constraint $C$ has the form
\eq{\label{eq:WKB-0}
\frac{\kappa}{2}G^{ab}(Q)\frac{\partial\W}{\partial Q^a}\frac{\partial\W}{\partial Q^b}+\frac{1}{\kappa}V(Q)+\Hs\left(Q;\frac{\partial \W}{\partial q},q\right) = 0 \ ,
}
where a summation over repeated indices is implied, $\kappa$ is a coupling constant (e.g., $\kappa = 4\pi G/3$ or $\kappa =1/c^2$), and $\W$ is the on-shell action. We can look for a Wentzel-Kramers-Brillouin (WKB) perturbative solution of~\eref{WKB-0} by formally expanding the on-shell action as follows:
\eq{\label{eq:WKB}
\W(Q,q) = \frac{1}{\kappa}\sum_{n = 0}^{\infty}\W_n(Q,q)\kappa^n =: \frac{1}{\kappa}\W_0(Q)+\mathtt{S}(Q;q) \ .
}
It is straightforward to verify that, at the lowest-oder, the solution satisfies the HJ constraint solely for the $Q$ variables, without any interaction with the $q$ fields,
\eq{
\frac{1}{2}G^{ab}(Q)\frac{\partial\W_0}{\partial Q^a}\frac{\partial\W_0}{\partial Q^b}+V(Q) = 0 \ .
}
This is the no-coupling limit. Its solution induces a foliation on the $Q$-configuration space, $Q\mapsto x(Q) = (t(Q), x^i(Q))$, where $t(Q)$ satisfies {\cite{MS-1,Kuchar}}
\eq{\label{eq:WKB-t}
G^{ab}\del{\W_0}{Q^a}\del{t}{Q^b} = 1 \ ;
}
i.e., it is the proper time in the no-coupling limit. Higher orders in the coupling constant $\kappa$ are encoded in the $\mathtt{S}(Q;q)$ function [cf.~\eref{WKB}], which obeys the ``corrected'' time-dependent HJ equation
\eq{\label{eq:HJcorr}
-\del{\mathtt{S}}{t} = \Hs-\frac{\kappa}{4V}\Hs^2+\frac{\kappa}{2}g^{ij}\del{\mathtt{S}}{x^i}\del{\mathtt{S}}{x^j}+\mathcal{O}(\kappa^2) \ .
}
Equation~\eref{HJcorr} implies that the no-coupling proper-time $t$ is the time that orders the evolution of the $(q,x)$ variables. It is a choice of gauge that is thus ``preferred'' by this weak-coupling expansion. The associated FP determinant is
\eq{\label{eq:FP-WKB}
|\Delta| = 1+\kappa\frac{\Hs}{2V}+\Ob(\kappa^2) \ .
}

\subsection{Weak-coupling expansion of the WDW equation. Unitarity}
The WDW equation associated with the classical constraint~\eref{WKB-0} reads
\eq{\label{eq:WDW-WKB}
\hat{C}\Psi = -\frac{\kappa}{2}\nabla^2\Psi+\frac{1}{\kappa}V(Q)\Psi+\hat{\Hs}\Psi = 0 \ ,
}
and, in analogy to the classical expansion~\eref{WKB}, we can solve~\eref{WDW-WKB} perturbatively via the ansatz
\eq{\label{eq:WKB-psi}
\Psi(Q,q) = \exp\left[ \frac{\I}{\kappa} \sum_{n = 0}^{\infty}\mathcal{W}_n(Q,q){\kappa^n}\right] =: \exp\squarebr{\I \frac{1}{\kappa} \W_0(Q,q)}\psi(Q;q) \ ,
}
which is frequently used in Born-Oppenheimer (BO) treatments of quantum cosmology {\cite{MS,MS-1}}. Just as in the classical case, we can adopt a foliation of the $Q$-configuration space defined from the no-coupling solution, and we find that $\psi(Q;q)$, which encodes the higher orders in $\kappa$, solves the corrected Schr\"{o}dinger equation
\eq{\label{eq:schro}
\I\del{\tilde{\psi}}{t} = \left[\hat{\Hs}-\frac{\kappa}{4V}\hat{\Hs}^2+\frac{\kappa}{2}\hat{\Pi}_ig^{ij}\hat{\Pi}_j^{\dagger}+\frac{\mathcal{Q}}{M}\right]\tilde{\psi} +\Ob(\kappa^2)\ ,
}
where $t$ is the no-coupling proper-time, $\mathcal{Q}$ is a quantum correction to the potential, and we have defined  
\eq{\label{eq:tilde-psi}
\tilde{\psi} := |2Vg|^{\frac14}\left(1+\frac{\kappa}{2V}\hat{\Hs}\right)\psi+\Ob(\kappa^2) \ .
}
Equation~\eref{schro} is relational in the sense that it describes the dynamics of the $(q,x)$ variables relative to the intrinsic time $t$. Moreover, it clearly leads to a unitary dynamics with respect to the inner product [cf. \eref{WKB-psi}, \eref{tilde-psi}]
\eq{\label{eq:IP-phys-WKB}
(\Psi_2|\Psi_1)&:= \int\D x\D q\ \tilde{\psi}_2^*(x,q)\tilde{\psi}_1(x,q)\\
&=\int\D Q\D q\sqrt{|2Vg|}\left|\del{x}{Q}\right|\ \Psi_2^*(x,q)\,\delta(t(Q)-s) \widehat{|\Delta|}\,\Psi_1(x,q)+\Ob(\kappa^2) \ ,
}
where
\eq{
\widehat{|\Delta|} := 1+\frac{\kappa}{2V}\hat{\Hs}
}
is the quantization of the (absolute value of the) classical FP determinant [cf.~\eref{FP-WKB}]. Thus, the inner product~\eref{IP-phys-WKB} is perturbatively defined and conserved, and it can be seen as a perturbative analogue of the physical inner product~\eref{phys-IP}. Moreover, the inner product~\eref{IP-phys-WKB} leads to a probability density
\eq{\label{eq:prob-WKB}
p_{\Psi} &:=\left.\frac{|\tilde{\psi}(x,q)|^2}{\left({\Psi|\Psi}\right)}\right|_{t = s}  \ , 
}
which refers to the measurement of the $(q,x)$ fields based on the condition that $t$ has (been observed to have) the value $s$. In this way, we can regard~\eref{prob-WKB} as conditional probabilities, in analogy to~\eref{relobs-cond}. In fact, relational observables can be perturbatively defined from their matrix elements by inserting operators in the integrand of~\eref{IP-phys-WKB} (see {\cite{MS-1,relobs-2}}).

\subsection{The early Universe}
This unitary, perturbative formalism was applied to a simple model of the early Universe in the first reference of {\cite{MS-1}}, which consisted of a de Sitter spacetime background [with scale factor $a = a_0\e^{\alpha}$ and cosmological constant $H_0^2/(2\kappa)$] with primordial perturbations (the Mukhanov-Sasaki variables denoted by $v$). The WDW constraint then has the form
\eq{
\left[\frac{\e^{-3\alpha}}{a_0}\left(\frac\kappa2\del{^2}{\alpha^2}+a_0^6\e^{6\alpha}\frac{H_0^2}{2\kappa}\right)+\frac{\e^{-\alpha}}{a_0}\hat{H}\right]\Psi(\alpha,v) = 0 \ , 
}
and it can be solved in perturbation theory following~\eref{WKB-psi}. The corrected Schr\"{o}dinger equation [cf.~\eref{schro}] reads
\eq{\label{eq:schro-uni}
\I\del{\tilde{\psi}}{\eta} = \left(\hat{H}-\kappa\frac{H_0^2\eta^4}{2}\hat{H}^2\right)\tilde{\psi}+\Ob(\kappa^2) \ ,
}
where the usual conformal time $\eta$ plays the role of the no-coupling proper time [cf.~\eref{WKB-t}]. From~\eref{schro-uni}, we can compute corrections to the primordial power spectra:
\eq{
\mathcal{P}_{S,T}(k)\simeq\mathcal{P}_{S,T;0}(k)\left\{1+\kappa H_0^2\left(\frac{k_{\star}}{k}\right)^3[2.85-2\log(-2k\eta)]\right\} \ .
}
This shows that this formalism can serve as a starting point for computing phenomenological effects in canonical quantum gravity within a relational, perturbative, and unitary framework. Possible further developments are mentioned in \cite{MS-1,relobs-2}.

\section{\label{sec:conclusions}Conclusions}
In this conference article, we have briefly reviewed two relational approaches to models that exhibit a local time reparametrization invariance. Both approaches are systematic, model independent, and they can be applied either in the classical or quantum theories.

The first approach is nonperturbative, and it is predicated on the possibility of solving the WDW equation exactly. The suggested symmetric construction of relational observables leads to the correct quantum evolution via a set of diffeomorphism-invariant Heisenberg equations of motion. If the observables are also self-adjoint (or have self-adjoint extensions), then this evolution can be shown to be unitary {\cite{relobs-11,relobs-2}}. Time is intrinsic and defined from the energy spectrum of the Hamiltonian constraint.

The second approach is perturbative and consists of a weak-coupling expansion that is frequently used in the BO approach to quantum cosmology. The main result is that this formalism is, in fact, a unitary gauge theory of local time reparametrizations. The choice of gauge is induced by the weak-coupling expansion, and leads to the definition of a particular time variable: the no-coupling proper time. The inner product is perturbatively defined and it is conserved (at least up to the next-to-leading order). This settles the long-standing issue of unitarity of corrections to the Schr\"{o}dinger equation in the BO approach. See {\cite{MS-1,relobs-2}} for a comparison and comments on different BO treatments. 

The two formalisms reviewed here can serve as starting points for a rigorous, relational understanding of dynamics in theories with local time reparametrization invariance, most notably quantum gravity. In particular, an adaptation of the weak-coupling expansion method to general slow-roll models could lead to unitary corrections to the usual expressions from which the power spectrum of the Cosmic Microwave Background radiation is derived. These corrections would constitute one avenue into the phenomenology of canonical quantum gravity. Moreover, the construction of quantum relational observables presented in Sec.~\ref{sec:qrelobs} together with the postulates given in Sec.~\ref{sec:postulates} could pave the way to a diffeomorphism-invariant description of the emergence of quantum mechanics. These are exciting prospects that ought to be considered in future research.

\ack{The author thanks Claus Kiefer, Manuel Kr\"{a}mer, David Brizuela, Branislav Nikoli\'{c}, J. Brian Pitts, Alexander Y. Kamenshchik, Alessandro Tronconi, and Giovanni Venturi for discussions. The author is also grateful to the Dipartimento di Fisica e Astronomia of the Universit\`{a} di Bologna as well as the I.N.F.N. Sezione di Bologna for financial support.}

\end{document}